# Coherent control of orbital wavefunctions in the quantum spin liquid $Tb_2Ti_2O_7$

R. Mankowsky[1], M. Müller[1], M. Sander[1], S. Zerdane[1], X. Liu[1], D. Babich[1], H. Ueda[1], Y. Deng[1], R. Winkler[2], B. Strudwick[1], M. Savoini[2], F. Giorgianni[1], S. L. Johnson[1,2], E. Pomjakushina[1], P. Beaud[1], T. Fennel[1], H.T. Lemke[1], U. Staub[1]

[1]*Paul Scherrer Institute, 5232 Villigen PSI, Switzerland*
[2]*Institute for Quantum Electronics, Physics Department, ETH Zürich, 8093 Zürich, Switzerland*

Resonant driving of electronic transitions with coherent laser sources creates quantum coherent superpositions of the involved electronic states. Most time-resolved studies have focused on gases or isolated subsystems embedded in insulating solids, aiming for applications in quantum information. Here, we demonstrate coherent control of orbital wavefunctions in pyrochlore $Tb_2Ti_2O_7$, which forms an interacting spin liquid ground state. We show that resonant excitation with a strong THz pulse creates a coherent superposition of the lowest energy Tb 4*f* states before the magnetic interactions eventually dephase them. The coherence manifests itself as a macroscopic oscillating magnetic dipole, which is detected by ultrafast resonant x-ray diffraction. The induced quantum coherence demonstrates coherent control of orbital wave functions, a new tool for the ultrafast manipulation and investigation of quantum materials.

The interplay between the orbital, spin and structural degrees of freedom is a driving factor of many key properties of solids, including high temperature superconductivity, colossal magneto resistance and metal-insulator transitions. In the last decades, great progress in controlling this interplay to obtain materials with novel functional properties has been made by techniques such as epitaxial strain engineering and the design of heterostructures (*1–3*). Adjustments in metal-insulator transition temperatures, modification of magnetic structures, or the creation of superconducting phases show the potential of these methods, which have focused in particular on transition metal oxides such as manganites (*4*) and nickelates (*5*). Driven by the development of new laser sources in recent years, the control over material properties was extended into the time domain by selective excitation of electronic, magnetic and structural degrees of freedom (*6–8*). While significant effort has been spent to coherently control collective excitations such as magnons (*9, 10*) and phonons (*11–14*), there are only few studies on coherent *orbital* excitations in correlated materials (*15–17*). Here, we demonstrate the resonant and coherent excitation of a purely orbital transition at THz frequency in a correlated rare earth quantum material.

The physics of these materials is defined by the rare earth ion's $4f$ valence electrons. They are partially shielded from the environment by the filled $5s^2$ and $5p^6$ shells, which reduces their interactions and leads to more localized states (*18*). This makes them ideal candidates to study orbital coherence in condensed matter. Enduring interest in the rare earth pyrochlore $Tb_2Ti_2O_7$ (TTO) (figure 1A) stems from its evasion of magnetic order and formation of a strongly correlated magnetic state below temperatures of the scale of the Curie-Weiss constant, 13 K (*19–24*), rather than long range magnetic order as originally anticipated (*25, 26*). Quantum fluctuations (*27*), structural distortions (*28*) or fluctuations (*21*), hybridization of magnetic and structural fluctuations (*19*), and quadrupolar degrees of freedom (*29*) have all been explored as mechanisms by which magnetic order may be suppressed while preserving the unprotected local moment of the non-Kramers doublet ground state of the $Tb^{3+}$ $^7F_6$ multiplet in the $D_{3d}$ symmetry crystal field (CF). The first excited crystal field level lies $\Delta E= h\nu_0=1.6$ meV ($\nu_0 = 0.4$ THz) above the ground state, as schematically shown on the right-hand side of figure 1A. Optical measurements show that on cooling TTO into the correlated state, an absorption peak appears at $\nu_0 = 0.44$, close to the lowest CF transition energy (*30*). The temperature dependence of its oscillator strength tracks the population difference between the ground and the first excited CF level, which are both doublets (*31*). Since the even symmetry ($E_g$) of the levels forbids an electric dipole excitation, the absorption is attributed to magnetic dipole coupling (*32, 33*). Here, we study the resonant excitation of this CF transition in the time domain. By measuring the magnetic and structural dynamics, we show that not only the population but also the quantum phase of the orbital wavefunction can be controlled.

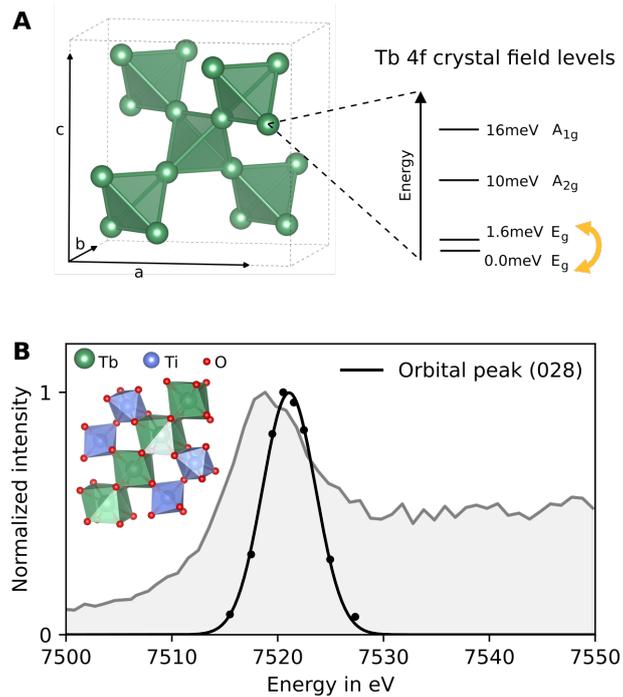

**Fig. 1: Cubic pyrochlore structure of $Tb_2Ti_2O_7$. A,** Lefthand side: corner-sharing tetrahedra, formed by the Tb ions (other ions are not shown for clarity). Righthand side: $Tb^{3+}$ $4f$ CF levels. The orange arrow indicates the transition, which was resonantly excited. **B,** X-ray absorption measured via total fluorescence yield across the Tb $L_3$-edge (grey in the background) and energy dependence of the orbital (028) diffraction peak (black). The inset shows the alternating orientation of the oxygen cages, which distinguish neighboring Tb sites (green). The crystal structures were visualized using the VESTA software (*34*).

In order to disentangle the orbital and magnetic dynamics from structural components of this transition, we employed time-resolved resonant and non-resonant x-ray diffraction using 50 fs x-ray pulses at the Bernina beamline of the SwissFEL Free Electron Laser (*35, 36*). Section 1 and Fig. S1 of the materials and methods section give more details on the experimental setup. For our experiments we use a single crystal of TTO with a (001) surface orientaion (sample EP3 from Ref. (*37*)). The orbital dynamics were studied by tuning the x-ray photon energy to the Tb $L_3$ edge. Figure 1B shows the photon energy dependence of the structurally forbidden (028) peak together with the x-ray absorption signal collected in total fluorescence yield mode. The intensity of the (028) peak shows a clear enhancement at 7522.5 eV, which originates from a periodic ordering of Tb orbitals caused by rotations of the oxygen cages surrounding them. Ionic motions were detected by following the diffracted intensity changes of the structurally allowed ($\bar{1}37$) diffraction peak.

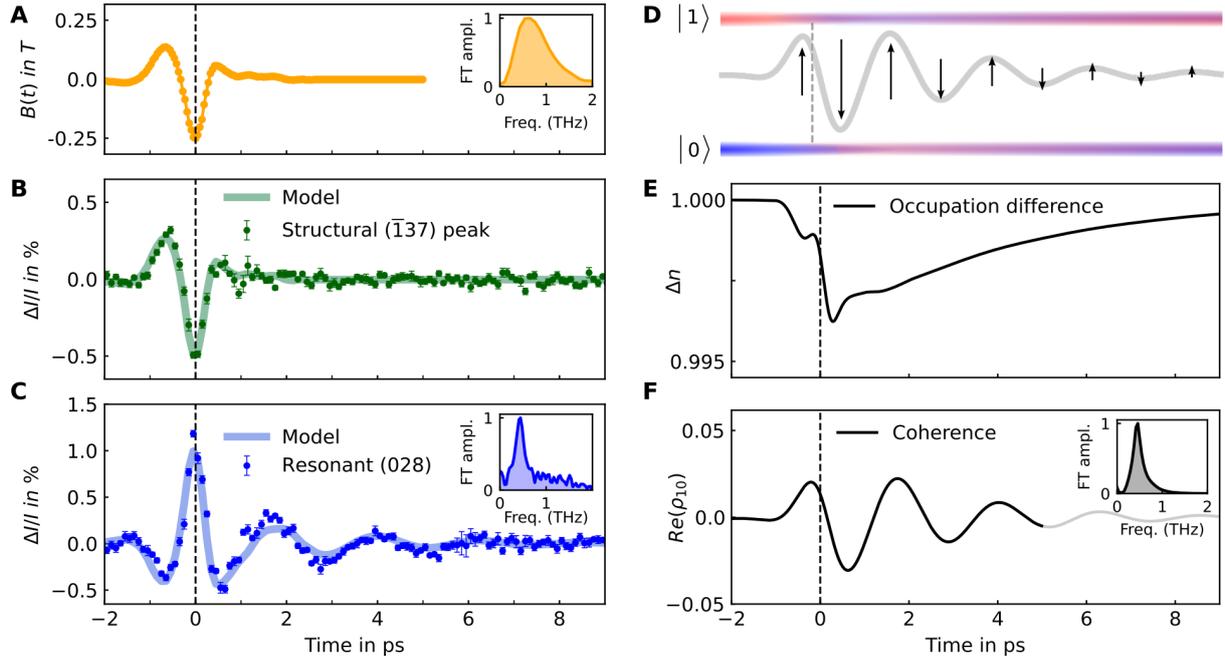

**Fig. 2: Time-dependent diffracted peak intensity at 5K sample temperature and simulation results**. **A**, Magnetic field of the THz pump pulses, obtained by electrooptic sampling (*38*). **B,C**, Change in diffracted intensity of the structural $(\bar{1}37)$ diffraction peak (green markers) and the resonant (028) diffraction peak (blue markers). Error bars reflect the 1σ (67%) confidence interval. Fourier transforms of the excitation pulse and the oscillatory component of the resonant peak are displayed in the insets. The experimental data are simulated (light solid lines) by fitting the coefficients of Equations (1) and (5) to the data. **D**, Simulation results. The resonant excitation induces a coherent superposition between ground state $|0\rangle$ (blue) and first excited Tb CF level $|1\rangle$ (red), which reduces their occupation difference $\Delta n$ by around 0.3% (**E**). The off-diagonal term of the density matrix $\rho_{10}$ (**F**) encapsulates the coherence of the CF levels, which manifests itself via a macroscopic oscillating magnetic dipole (grey line) due to the collective motion of the Tb spins (black arrows), which oscillate coherently with frequency $\nu_0 = 0.44$ THz.

The single cycle electromagnetic pulses used for the excitation reached peak electric and magnetic fields of 0.75 MV/cm and 0.25 T, respectively, with a central frequency of 0.6 THz as shown in figure 2A. The linearly polarized electric and magnetic field components were polarized along the sample [110] and [1$\bar{1}$0] directions, respectively. The same figure shows the temporal evolution of the change in diffracted intensity of the structural $(\bar{1}37)$ and the resonant orbital (028) diffraction peaks following excitation at 5 K sample temperature. We first discuss the time-resolved changes of the $(\bar{1}37)$ peak intensity (green, panel B), which is insensitive to induced magnetic or orbital dynamics. Since all optically active phonon modes of TTO have frequencies >2.7 THz (*30, 32*), significantly higher than the driving field, they are not resonantly excited. Under these conditions, the ions simply follow the THz pump electric field $E(t)$ in phase, like oscillators driven below their resonance frequency. Hence, the associated changes in diffraction intensity are proportional to the electric field $E(t)$, contributing to the change in diffracted intensity according to

$$\left(\frac{\Delta I}{I}\right)_{struc}(t) = a_s \cdot E(t), \tag{1}$$

where $a_s$ is a constant.

At early times, the diffraction intensity of the resonant peak (blue, panel C) shows similar behavior, albeit with opposite sign. This behavior is largely expected: the same ionic motion that is measured in the $(\bar{1}37)$ peak is likely to also influence the position, orientation and shape of the Tb orbitals due to changes of the CF potential. The time-resolved intensity change of the (028) peak, however, has an important additional feature: a coherently oscillating component that persists beyond the driving pulse. Despite the broad excitation bandwidth, the spectrum of this induced oscillatory component is sharply peaked at $\nu_0 = 0.44$ THz, which corresponds to the energy difference $\Delta E$ between the resonantly excited levels (see Fourier transforms in panels A and C of figure 2).

In order to better understand the origin of this coherent signal, we simulate the dynamics induced by the coherent excitation of the first Tb CF level at $\Delta E=1.6$ meV. Since further CF levels are separated by $\geq 10$ meV, they can be neglected. For a given ion, the direction of the transient magnetic field selects a preferential basis within the two doublets, coupling one singlet of the ground state doublet $|0\rangle$ to one partner singlet in the excited state $|1\rangle$. These pairs of ground and excited states form an ensemble of effective quantum two-level systems (TLS), each described by a wave function $|\psi\rangle = c_0|0\rangle + c_1|1\rangle$ (see supplementary text 1 for a detailed derivation). Given the moderate exchange and dipolar couplings between different Tb ions of ~ 1meV estimated from the Curie-Weiss temperature, the short time dynamics of the ensemble is well described by the density matrix formalism for single ions (*39*). The average state of an effective TLS is given by its density matrix $\rho$:

$$\rho = \begin{pmatrix} \overline{c_0^2} & \overline{c_1^* c_0} \\ \overline{c_0^* c_1} & \overline{c_1^2} \end{pmatrix}, \quad \rho_{10} = \rho_{01}^*. \quad (2)$$

The diagonal elements $\rho_{00} = \overline{|c_0^2|}$ and $\rho_{11} = \overline{|c_1^2|} = \frac{1}{2} - \overline{|c_0^2|}$ represent the occupation of the ground and the excited state, respectively, while the off-diagonal elements $\rho_{10}$ and $\rho_{01}$ capture their phase coherence.

When the atoms are exposed to the magnetic field $B(t)$ of a light pulse, the dynamics of the occupation difference $\Delta n = \rho_{00} - \rho_{11}$ and the induced coherence $\rho_{10}$ evolve according to the coupled equations of motion:

$$\frac{d\Delta n(t)}{dt} = i\frac{2dg_J\mu_B B(t)}{\hbar}\left(\rho_{10}(t) - \rho_{10}^*(t)\right) - \frac{\Delta n(t) - \Delta n_{eq}}{\tau}; \quad \Delta n(t=-\infty) = \Delta n_{eq}. \quad (3)$$

$$\frac{d\rho_{10}(t)}{dt} = -i\omega_0 \rho_{10}(t) + i\frac{dg_J\mu_B B(t)}{\hbar}\Delta n(t) - \frac{\rho_{10}(t)}{\tau_c}; \quad \rho_{10}(t=-\infty) = 0. \quad (4)$$

Here, $\omega_0 = 2\pi\nu_0$ is the resonance angular frequency and the matrix element $d \approx 3$ can be viewed as the effective magnetic moment of the two-level system. The prefactors $g_J = 3/2$ and $\mu_B$ denote the Landé factor of the Tb $^7F_6$ multiplet and the Bohr magneton. The first term in Eq. (3) describes the coherent build-up of an excited population with instantaneous Rabi frequency $\Omega_R(t) = dg_J\mu_B B(t)$. The second term describes relaxation back to the equilibrium value $\Delta n_{eq}$, with the lifetime of the excited state $\tau$. The first two terms in Eq. (4) describe the unitary evolution of the coherence $\rho_{10}$, while the last term captures its dephasing over a decoherence time $\tau_c$ with $\tau_c \leq \tau$.

Since the (028) diffraction peak has contributions from transitions into all unoccupied Tb 4$f$ crystal field levels, our experiment is insensitive to moderate changes in the occupation of only the lowest CF levels. In contrast, the coherence manifests itself as a macroscopic magnetic dipole moment $\overline{\langle\mu\rangle}(t) = tr(\rho(t)\mu) = \mu(\rho_{10}(t) + \rho_{10}^*(t))$, which can be detected with high sensitivity. Due to the inequivalent quantization axes of the Tb ions defined by their orbital orientations, this transient magnetic ordering of the spins constitutes an additional contribution to the x-ray scattering of the (028) diffraction peak at resonance (see Methods section 2). This dynamic magnetic contribution interferes with the static orbital contribution, which strongly enhances its detection sensitivity. Therefore, the total change in diffracted intensity of the (028) peak contains an additional magnetic contribution proportional to $Re(\rho_{10})(t)$ driven by the magnetic field $B(t)$ of the THz pulse. This results in an overall change to the (028) diffraction peak intensity in resonance of

$$\left(\frac{\Delta I}{I}\right)_{res}(t) = a_{s,o} \cdot E(t) + a_m \cdot Re(\rho_{10})(t), \qquad (5)$$

where $a_{s,o}$ and $a_m$ are constants.

We now use equations (1) and (5) to reproduce the experimental data shown in figure 2B and 2C, respectively. The electric $E(t)$ and magnetic fields $B(t)$ of the THz pulse are measured via electro-optic sampling. The measured magnetic field is used to obtain $\rho_{10}(t)$ via numerical solution of equations (3) and (4). The parameters $a_s, a_{s\_o}, a_m, \nu_0, \tau_c$ are adjusted to best fit the data (see table S2 for a summary of the obtained values). Since the induced magnetization is sensitive only to the coherence time $\tau_c$, the lifetime of the excited state $\tau$ cannot be used as fitting parameter and has been fixed instead to 4 ps estimated from the Tb-Tb hopping time given by their interaction energy of 1 meV. The resulting change in diffracted intensity is displayed in figure 2B and 2C as light lines, which agree well with the experimental data. The solution of the differential equations (3, 4) is visualized in panels D-F of figure 2.

Following the excitation, the occupation difference (black curve) decreases, as 0.15% of the Tb ions are excited from the ground state to the higher energy level. The coherent superposition of the two CF levels is reflected in coherent oscillations of the off-diagonal term $\rho_{10}$ of the density matrix (figure 2F). These oscillations with frequency $\nu_0 = 0.44$ THz persist on the order of the coherence time $\tau_c$ beyond the end of the exciting pulse. The obtained coherence time $\tau_c$ of $2.4 \pm 0.4$ ps is of the same order as the timescale set by Tb-Tb exchange and dipolar couplings, indicating that the decoherence is dominated by the intrinsic lifetime of the excited state.

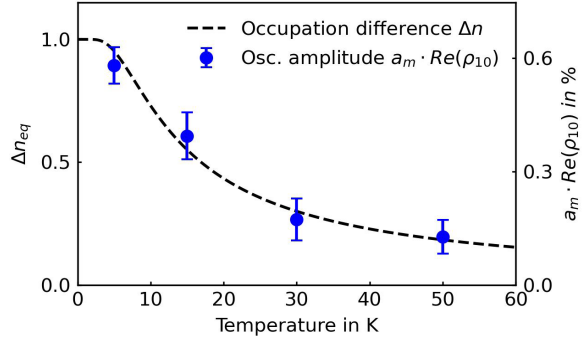

**Fig. 3: Temperature dependence of the induced magnetic dipole.** The data points show the amplitude of the oscillatory magnetic component observed in the diffraction intensity change of the resonant peak. They have been extracted from fits of the model coefficients (Equations (1) and (5)) to the data (right vertical axis). Error bars reflect the $1\sigma$ (67%) confidence interval. The dashed black line visualizes the occupation difference $\Delta n_{eq}(T)$ of the two levels calculated from Boltzmann statistics (left vertical axis). See figure S2 for the individual time-traces of the orbital and structural peak.

Finally, we compare the temperature dependence of the oscillation amplitude, i.e. the coherence of the ensemble, with the predictions of the model. The induced coherence depends linearly on the magnetic field and the occupation difference $\Delta n_{eq}$ between the two CF states before excitation (see Eq. 3 and Supplementary Text Eq. 22). The temperature dependence of the amplitude obtained by the fits to the data shown in figure 3 follows Boltzmann statistics $\Delta n_{eq}(T) = tanh(\Delta E/2k_B T)$ as predicted, which further validates our model. Signatures of induced orbital coherence are found up to 50 K with a constant coherence time of 2.4 ps. As the coherence time is limited by Tb-Tb interactions, it can be substantially increased by diluting rare earth ions in a nonmagnetic parent compound (*17*). Our results demonstrate the coherent control of orbital wavefunctions in a dense, correlated quantum material at temperatures reaching 50 K, supporting recent proposals to use ions with localized 4*f* shells in quantum computing schemes (*40*).

More generally the coherent control of orbitals constitutes a new tool for the ultrafast control of materials to manipulate magnetic properties on fs-ps timescales with elemental selectivity. The excitation with linearly polarized light – as used here – induces an oscillatory magnetization on atomic sites selected by the crystal field transition. However, upon changing polarization and direction of the THz pulse, the driven 4-level systems will split differently into two effective TLS resulting in different responses. In particular, circularly polarized light would induce non-oscillatory, quasi-static magnetization, which rises within the pulse duration and survives on the order of the life time $\tau$ beyond the pulse (see Supplementary Text section 1.9). This ultrafast magnetic pulse reaches moments of the order of a Bohr magneton per site, creating internal fields of around 20 mT in $Tb_2Ti_2O_7$. So far most efforts to achieve ultrafast control of materials has centered on the concepts of photodoping or the excitation of collective excitations such as phonons, magnons or electromagnons. We envision the coherent control of orbitals demonstrated here to become part of this toolbox to explore out of equilibrium phases of materials and to tune collective properties in analogy to the equilibrium control of orbital polarization and occupation by strain, fields and pressure. Further developments of this technique might open new ways to probe correlations and entanglement among ions in close proximity.

**Acknowledgements:**

We acknowledge the Paul Scherrer Institut, Villigen, Switzerland, for provision of beamtime at the Bernina beamline of SwissFEL as well as at the Materials Science beamline of the Swiss Light Source (SLS). We thank B. Pedrini for his excellent support.We are grateful for fruitful discussions with R. Ballou, S. deBrion and with A. Herzig.

R.M. and H.U. acknowledge funding from the National Centers of Competence in Research in Molecular Ultrafast Science and Technology (NCCR MUST-No. 51NF40-183615) from the Swiss National Science Foundation

H.U. also acknowledges funding from the European Union's Horizon 2020 innovation program under the Marie Skłodowska-Curie Grant Agreement No. 801459 – FP-RESOMUS.

**References:**

1.  D. G. Schlom, L.-Q. Chen, C.-B. Eom, K. M. Rabe, S. K. Streiffer, J.-M. Triscone, Strain Tuning of Ferroelectric Thin Films. *Annu. Rev. Mater. Res.* **37**, 589–626 (2007).

2.  P. Zubko, S. Gariglio, M. Gabay, P. Ghosez, J.-M. Triscone, Interface Physics in Complex Oxide Heterostructures. *Annu. Rev. Condens. Matter Phys.* **2**, 141–165 (2011).

3.  H. Y. Hwang, Y. Iwasa, M. Kawasaki, B. Keimer, N. Nagaosa, Y. Tokura, Emergent phenomena at oxide interfaces. *Nat. Mater.* **11**, 103–113 (2012).

4.  D. Yi, N. Lu, X. Chen, S. Shen, P. Yu, Engineering magnetism at functional oxides interfaces: manganites and beyond. *J. Phys. Condens. Matter*. **29**, 443004 (2017).

5.  S. Catalano, M. Gibert, J. Fowlie, J. Íñiguez, J.-M. Triscone, J. Kreisel, Rare-earth nickelates $R$NiO$_3$ : thin films and heterostructures. *Rep. Prog. Phys.* **81**, 046501 (2018).

6.  A. D. Caviglia, R. Scherwitzl, P. Popovich, W. Hu, H. Bromberger, R. Singla, M. Mitrano, M. C. Hoffmann, S. Kaiser, P. Zubko, S. Gariglio, J.-M. Triscone, M. Först, A. Cavalleri, Ultrafast Strain Engineering in Complex Oxide Heterostructures. *Phys. Rev. Lett.* **108**, 136801 (2012).

7.  P. Beaud, A. Caviezel, S. O. Mariager, L. Rettig, G. Ingold, C. Dornes, S.-W. Huang, J. A. Johnson, M. Radovic, T. Huber, T. Kubacka, A. Ferrer, H. T. Lemke, M. Chollet, D. Zhu, J. M. Glownia, M. Sikorski, A. Robert, H. Wadati, M. Nakamura, M. Kawasaki, Y. Tokura, S. L. Johnson, U. Staub, A time-dependent order parameter for ultrafast photoinduced phase transitions. *Nat. Mater.* **13**, 923–927 (2014).

8.  S. Wall, S. Yang, L. Vidas, M. Chollet, J. M. Glownia, M. Kozina, T. Katayama, T. Henighan, M. Jiang, T. A. Miller, D. A. Reis, L. A. Boatner, O. Delaire, M. Trigo, Ultrafast disordering of vanadium dimers in photoexcited VO2. *Science*. **362**, 572–576 (2018).

9.  P. Němec, M. Fiebig, T. Kampfrath, A. V. Kimel, Antiferromagnetic opto-spintronics. *Nat. Phys.* **14**, 229–241 (2018).

10. T. Kubacka, J. A. Johnson, M. C. Hoffmann, C. Vicario, S. de Jong, P. Beaud, S. Grübel, S.-W. Huang, L. Huber, L. Patthey, Y.-D. Chuang, J. J. Turner, G. L. Dakovski, W.-S. Lee, M. P. Minitti, W. Schlotter, R. G. Moore, C. P. Hauri, S. M. Koohpayeh, V. Scagnoli, G. Ingold, S. L. Johnson, U. Staub, Large-Amplitude Spin Dynamics Driven by a THz Pulse in Resonance with an Electromagnon. *Science*. **343**, 1333–1336 (2014).

11. R. Mankowsky, M. Först, A. Cavalleri, Non-equilibrium control of complex solids by nonlinear phononics. *Rep. Prog. Phys.* **79**, 064503 (2016).

12. S. Maehrlein, A. Paarmann, M. Wolf, T. Kampfrath, Terahertz Sum-Frequency Excitation of a Raman-Active Phonon. *Phys. Rev. Lett.* **119**, 127402 (2017).

13. A. Stupakiewicz, C. S. Davies, K. Szerenos, D. Afanasiev, K. S. Rabinovich, A. V. Boris, A. Caviglia, A. V. Kimel, A. Kirilyuk, Ultrafast phononic switching of magnetization. *Nat. Phys.* **17**, 489–492 (2021).

14. R. Mankowsky, A. Subedi, M. Först, S. O. Mariager, M. Chollet, H. T. Lemke, J. S. Robinson, J. M. Glownia, M. P. Minitti, A. Frano, M. Fechner, N. A. Spaldin, T. Loew, B. Keimer, A. Georges, A. Cavalleri, Nonlinear lattice dynamics as a basis for enhanced superconductivity in YBa 2 Cu 3 O 6.5. *Nature*. **516**, 71–73 (2014).

15. D. Polli, M. Rini, S. Wall, R. W. Schoenlein, Y. Tomioka, Y. Tokura, G. Cerullo, A. Cavalleri, Coherent orbital waves in the photo-induced insulator–metal dynamics of a magnetoresistive manganite. *Nat. Mater.* **6**, 643–647 (2007).

16. S. Wall, D. Brida, S. R. Clark, H. P. Ehrke, D. Jaksch, A. Ardavan, S. Bonora, H. Uemura, Y. Takahashi, T. Hasegawa, H. Okamoto, G. Cerullo, A. Cavalleri, Quantum interference between charge excitation paths in a solid-state Mott insulator. *Nat. Phys.* **7**, 114–118 (2011).

17. A. Beckert, M. Grimm, N. Wili, R. Tschaggelar, G. Jeschke, G. Matmon, S. Gerber, M. Müller, G. Aeppli, Emergence of highly coherent quantum subsystems of a noisy and dense spin system (2022), , doi:10.48550/arXiv.2210.01024.

18. C. W. Thiel, T. Böttger, R. L. Cone, Rare-earth-doped materials for applications in quantum information storage and signal processing. *J. Lumin.* **131**, 353–361 (2011).

19. T. Fennell, M. Kenzelmann, B. Roessli, H. Mutka, J. Ollivier, M. Ruminy, U. Stuhr, O. Zaharko, L. Bovo, A. Cervellino, M. K. Haas, R. J. Cava, Magnetoelastic Excitations in the Pyrochlore Spin Liquid $Tb_2Ti_2O_7$. *Phys. Rev. Lett.* **112**, 017203 (2014).

20. A. A. Turrini, M. Ruminy, F. Bourdarot, U. Stuhr, J. S. White, G. Tucker, M. Skoulatos, M. Núñez-Valdez, T. Fennell, Magnetic-field control of magnetoelastic coupling in the rare-earth pyrochlore $Tb_2Ti_2O_7$. *Phys. Rev. B*. **104**, 224403 (2021).

21. J. P. C. Ruff, B. D. Gaulin, J. P. Castellan, K. C. Rule, J. P. Clancy, J. Rodriguez, H. A. Dabkowska, Structural Fluctuations in the Spin-Liquid State of $Tb_2Ti_2O_7$. *Phys. Rev. Lett.* **99**, 237202 (2007).

22. J. P. C. Ruff, Z. Islam, J. P. Clancy, K. A. Ross, H. Nojiri, Y. H. Matsuda, H. A. Dabkowska, A. D. Dabkowski, B. D. Gaulin, Magnetoelastics of a Spin Liquid: X-Ray

Diffraction Studies of $Tb_2Ti_2O_7$ in Pulsed Magnetic Fields. *Phys. Rev. Lett.* **105**, 077203 (2010).

23. M. Ruminy, S. Guitteny, J. Robert, L.-P. Regnault, M. Boehm, P. Steffens, H. Mutka, J. Ollivier, U. Stuhr, J. S. White, B. Roessli, L. Bovo, C. Decorse, M. K. Haas, R. J. Cava, I. Mirebeau, M. Kenzelmann, S. Petit, T. Fennell, Magnetoelastic excitation spectrum in the rare-earth pyrochlore $Tb_2Ti_2O_7$. *Phys. Rev. B*. **99**, 224431 (2019).

24. M. J. P. Gingras, B. C. den Hertog, M. Faucher, J. S. Gardner, S. R. Dunsiger, L. J. Chang, B. D. Gaulin, N. P. Raju, J. E. Greedan, Thermodynamic and single-ion properties of $Tb^{3+}$ within the collective paramagnetic-spin liquid state of the frustrated pyrochlore antiferromagnet $Tb_2Ti_2O_7$. *Phys. Rev. B*. **62**, 6496–6511 (2000).

25. J. G. Rau, M. J. P. Gingras, Frustrated Quantum Rare-Earth Pyrochlores. *Annu. Rev. Condens. Matter Phys.* **10**, 357–386 (2019).

26. J. S. Gardner, M. J. P. Gingras, J. E. Greedan, Magnetic pyrochlore oxides. *Rev Mod Phys*. **82**, 53 (2010).

27. H. R. Molavian, M. J. P. Gingras, B. Canals, Dynamically Induced Frustration as a Route to a Quantum Spin Ice State in $Tb_2Ti_2O_7$ via Virtual Crystal Field Excitations and Quantum Many-Body Effects. *Phys. Rev. Lett.* **98**, 157204 (2007).

28. P. Bonville, I. Mirebeau, A. Gukasov, S. Petit, J. Robert, Tetragonal distortion yielding a two-singlet spin liquid in pyrochlore $Tb_2Ti_2O_7$. *Phys. Rev. B*. **84**, 184409 (2011).

29. H. Takatsu, S. Onoda, S. Kittaka, A. Kasahara, Y. Kono, T. Sakakibara, Y. Kato, B. Fåk, J. Ollivier, J. W. Lynn, T. Taniguchi, M. Wakita, H. Kadowaki, Quadrupole Order in the Frustrated Pyrochlore $Tb_{2+x}Ti_{2-x}O_{7+y}$. *Phys. Rev. Lett.* **116**, 217201 (2016).

30. T. T. A. Lummen, I. P. Handayani, M. C. Donker, D. Fausti, G. Dhalenne, P. Berthet, A. Revcolevschi, P. H. M. van Loosdrecht, Phonon and crystal field excitations in geometrically frustrated rare earth titanates. *Phys. Rev. B*. **77**, 214310 (2008).

31. M. Ruminy, E. Pomjakushina, K. Iida, K. Kamazawa, D. T. Adroja, U. Stuhr, T. Fennell, Crystal-field parameters of the rare-earth pyrochlores $R_2Ti_2O_7$ (R = Tb, Dy, and Ho). *Phys. Rev. B*. **94**, 024430 (2016).

32. E. Constable, R. Ballou, J. Robert, C. Decorse, J.-B. Brubach, P. Roy, E. Lhotel, L. Del-Rey, V. Simonet, S. Petit, S. deBrion, Double vibronic process in the quantum spin ice candidate $Tb_2Ti_2O_7$ revealed by terahertz spectroscopy. *Phys. Rev. B*. **95**, 020415 (2017).

33. K. Amelin, Y. Alexanian, U. Nagel, T. Rõõm, J. Robert, J. Debray, V. Simonet, C. Decorse, Z. Wang, R. Ballou, E. Constable, S. de Brion, Terahertz magneto-optical investigation of quadrupolar spin-lattice effects in magnetically frustrated $Tb_2Ti_2O_7$. *Phys. Rev. B*. **102**, 134428 (2020).

34. K. Momma, F. Izumi, VESTA 3 for three-dimensional visualization of crystal, volumetric and morphology data. *J. Appl. Crystallogr.* **44**, 1272–1276 (2011).

35. G. Ingold, R. Abela, C. Arrell, P. Beaud, P. Böhler, M. Cammarata, Y. Deng, C. Erny, V. Esposito, U. Flechsig, R. Follath, C. Hauri, S. Johnson, P. Juranic, G. F. Mancini, R.

Mankowsky, A. Mozzanica, R. A. Oggenfuss, B. D. Patterson, L. Patthey, B. Pedrini, J. Rittmann, L. Sala, M. Savoini, C. Svetina, T. Zamofing, S. Zerdane, H. T. Lemke, Experimental station Bernina at SwissFEL: condensed matter physics on femtosecond time scales investigated by X-ray diffraction and spectroscopic methods. *J. Synchrotron Radiat.* **26**, 874–886 (2019).

36. E. Prat, R. Abela, M. Aiba, A. Alarcon, J. Alex, Y. Arbelo, C. Arrell, V. Arsov, C. Bacellar, C. Beard, P. Beaud, S. Bettoni, R. Biffiger, M. Bopp, H.-H. Braun, M. Calvi, A. Cassar, T. Celcer, M. Chergui, P. Chevtsov, C. Cirelli, A. Citterio, P. Craievich, M. C. Divall, A. Dax, M. Dehler, Y. Deng, A. Dietrich, P. Dijkstal, R. Dinapoli, S. Dordevic, S. Ebner, D. Engeler, C. Erny, V. Esposito, E. Ferrari, U. Flechsig, R. Follath, F. Frei, R. Ganter, T. Garvey, Z. Geng, A. Gobbo, C. Gough, A. Hauff, C. P. Hauri, N. Hiller, S. Hunziker, M. Huppert, G. Ingold, R. Ischebeck, M. Janousch, P. J. M. Johnson, S. L. Johnson, P. Juranić, M. Jurcevic, M. Kaiser, R. Kalt, B. Keil, D. Kiselev, C. Kittel, G. Knopp, W. Koprek, M. Laznovsky, H. T. Lemke, D. L. Sancho, F. Löhl, A. Malyzhenkov, G. F. Mancini, R. Mankowsky, F. Marcellini, G. Marinkovic, I. Martiel, F. Märki, C. J. Milne, A. Mozzanica, K. Nass, G. L. Orlandi, C. O. Loch, M. Paraliev, B. Patterson, L. Patthey, B. Pedrini, M. Pedrozzi, C. Pradervand, P. Radi, J.-Y. Raguin, S. Redford, J. Rehanek, S. Reiche, L. Rivkin, A. Romann, L. Sala, M. Sander, T. Schietinger, T. Schilcher, V. Schlott, T. Schmidt, M. Seidel, M. Stadler, L. Stingelin, C. Svetina, D. M. Treyer, A. Trisorio, C. Vicario, D. Voulot, A. Wrulich, S. Zerdane, E. Zimoch, A compact and cost-effective hard X-ray free-electron laser driven by a high-brightness and low-energy electron beam. *Nat. Photonics.* **14**, 748–754 (2020).

37. M. Ruminy, L. Bovo, E. Pomjakushina, M. K. Haas, U. Stuhr, A. Cervellino, R. J. Cava, M. Kenzelmann, T. Fennell, Sample independence of magnetoelastic excitations in the rare-earth pyrochlore $Tb_2Ti_2O_7$. *Phys. Rev. B.* **93**, 144407 (2016).

38. R. Mankowsky, M. Sander, S. Zerdane, J. Vonka, M. Bartkowiak, Y. Deng, R. Winkler, F. Giorgianni, G. Matmon, S. Gerber, P. Beaud, H. T. Lemke, New insights into correlated materials in the time domain—combining far-infrared excitation with x-ray probes at cryogenic temperatures. *J. Phys. Condens. Matter.* **33**, 374001 (2021).

39. U. Fano, Description of States in Quantum Mechanics by Density Matrix and Operator Techniques. *Rev. Mod. Phys.* **29**, 74–93 (1957).

40. M. Grimm, A. Beckert, G. Aeppli, M. Müller, Universal Quantum Computing Using Electronuclear Wavefunctions of Rare-Earth Ions. *PRX Quantum.* **2**, 010312 (2021).

# Materials and Methods for

# Coherent control of orbital wavefunctions in the quantum spin liquid Tb$_2$Ti$_2$O$_7$

## 1 Experimental setup and geometry

The experimental geometry is shown in figure S1. The THz pulses (yellow) were generated outside the chamber by optical rectification of 7.4 mJ, 100 fs, 800 nm pulses in LiNbO$_3$. They were guided through an 8mm thick ZEONEX window into the vacuum chamber and focused by a 2inch diameter, 2inch focal distance parabolic mirror onto the sample. Their field strength and spectral content was characterized in vacuum by electrooptic sampling with 100fs 800nm pulses (red) in a 100um thick GaP crystal mounted next to the sample. The FWHM diameter of the 800nm sampling pulses on the GaP was 90um, slightly larger than the 75um diameter x-ray pulses (dark blue), which were guided through a hole in the parabolic mirror onto the sample. The energy of the 50fs x-ray pulses was defined with a double crystal Si (111) monochromator to a bandwidth of 1eV. In diffraction condition of the (028) and ($\bar{1}$37) peaks, the incidence angle of the p-polarized THz pulses was 58.7° with respect to the sample surface. Due to the large refractive index of $n > 7.3$ within the spectral content of the THz pulses (extracted from reference *(30)*), they were strongly refracted into the sample, almost propagating along the [001] sample surface normal direction. The electric and magnetic field components were polarized along the [110] and [1$\bar{1}$0] directions. The sample was cooled with a helium flow cryostat to a base temperature of 5 K.

## 2 Orbital and magnetic contributions to the diffraction intensity of the resonant (028) diffraction peak

### 2.1. Static orbital contribution:
The site symmetry of the Tb ion at Wyckoff position 16c of the space group Fd-3m (227) is $-3m$ with a multiplicity of 16. The symmetry operations generate 4 unique Tb ions at positions Tb1 @ (0, 0, 0)$_{cubic}$, Tb2 @ (3/4, 1/4, 1/2)$_{cubic}$, Tb3 @ (1/4, 1/2, 3/4)$_{cubic}$, and Tb4 @ (1/2, 3/4, 1/4)$_{cubic}$. The remaining Tb ions in the unit cell are generated by translations of the unique Tb atoms by (0, 0, 0), (0, 1/2, 1/2), (1/2, 0, 1/2), and (1/2, 1/2, 0). Because of the local -3m symmetry, the only quadrupole moment that can contribute to the (028) equilibrium intensity is $Q_{3z^2-r^2} = \frac{1}{2}(2f_{zz} - f_{xx} - f_{yy})$, where *z* points along the local $C_3$ symmetry axis.

### 2.2. Dynamic magnetic contribution:
In the following, we evaluate the contribution of magnetic scattering to the (028) diffraction intensity due to transient magnetic moments $\vec{m_i}$ induced by the resonant excitation of the first CFL on the 4 unique Tb sites denoted by the index *i*. Following equation (35), the induced magnetic moment can be evaluated by projecting the [1-10] polarized B-field onto the local quantization axis $xyz_{local,i}$. Using the local Cartesian coordinate with *z* parallel to the $C_3$ symmetry axis for Tb1 @ (0, 0, 0)$_{cubic}$, we obtain $x$ // [1-10], $y$ // [11-2], and $z$ // [111]. Table S1 summarizes the positions, quantization axis as well as the induced magnetic moments of all 4 unique Tb ions. The magnetic form factor $\vec{F}_m(t) = \sum_i \vec{m_i}(t) e^{2\pi i(hx+ky+lz)}$ among the four Tb atoms can then be evaluated as:

$$\vec{F}_m(t) \approx -3.4 M(t) \cdot (1, -1, 0),$$

with $M(t) = (g_J \mu_B)^2 sin(\omega_0 t) \int_{-\infty}^{t} dt' B_0(t') e^{-(t-t')/\tau_c}$ following from equation (32). The total magnetic scattering factor of a single unit cell is just a multiplication of the above form factor by the factor of four. The incident x-ray pulses were horizontally polarized along $\varepsilon_i = [1 - 1 0]$ and the (028) peak was measured without polarization analysis. Since magnetic scattering is given by the projections of $\vec{F}_m // (1, -1, 0)$ onto the propagations of the incident and scattered x-rays as well as its cross product, there is a finite contribution of magnetic scattering to the (028) reflection, which can interfere with the orbital scattering given by $Q_{3z^2-r^2}$ and induces a linear modulation of the orbital peak intensity.

# Supplementary Materials for

# Coherent control of orbital wavefunctions in the quantum spin liquid Tb$_2$Ti$_2$O$_7$

## 1 Dynamics of the Tb 4-level system

### 1.1 Wave functions of the crystal field levels

The point group of the Tb ions within the pyrochlore lattice is $C_3$. We refer to the 3-fold rotation axis as the local z-axis. This symmetry implies that $J_z$ modulo 3 ($J_z mod 3$) is a good quantum number, that is, the crystal field levels within the ground state multiplet are superpositions of levels with values $J_z$ whose differences are multiples of 3. Non-degenerate CF levels are formed by the eigenstates with $J_z mod 3 = 0$. The eigenstates with $J_z mod 3 = 1,2$ form symmetry protected multiplets with their time reversed counterparts (taking values of $J_z$ mod 3 with opposite sign). The two lowest-lying CF eigenstates of Tb in Tb$_2$Ti$_2$O$_7$ form two such doublets, which we denote by $G$ (ground) and $E$ (excited). We further refer to the states with fixed $J_z mod 3$ as ↑/↓, according to the sign of the expectation value $\langle J_z \rangle$ of the magnetic moment along z.

The diagonalization of the CF Hamiltonian yields the explicit wavefunctions (using the shorthand notation $|m\rangle = |j=6, m\rangle$) using the coefficients tabulated in table IX of reference (31).

$$|G \uparrow\rangle = a_1|4\rangle - b_1|1\rangle + c_1|-2\rangle + d_1|-5\rangle \quad (1)$$
$$|G \downarrow\rangle = T|G \uparrow\rangle = a_1|-4\rangle + b_1|-1\rangle + c_1|2\rangle - d_1|5\rangle \quad (2)$$

with $a_1 = 0.9116$, $b_1 = 0.1186$, $c_1 = 0.1765$, $d_1 = 03519$, whereby the time reversal operator acts as $T|m\rangle = (-1)^m|-m\rangle$. The magnetic moment of the ground state doublet is $m_G = \langle G \uparrow |J_z|G \uparrow\rangle = -\langle G \downarrow |J_z|G \downarrow\rangle = 2.657$.

For the excited doublet one finds
$$|E \uparrow\rangle = a_2|5\rangle - b_2|2\rangle + c_2|-1\rangle + d_2|-4\rangle \quad (3)$$
$$|E \downarrow\rangle = T|E \uparrow\rangle = -a_2|-5\rangle - b_2|-2\rangle - c_2|1\rangle + d_2|4\rangle \quad (4)$$

with $a_2 = 0.8910$, $b_2 = 0.2320$, $c_2 = 0.1085$, $d_2 = 0.3748$, having a larger magnetic moment $m_E = \langle E \uparrow |J_z|E \uparrow\rangle = -\langle E \downarrow |J_z|E \downarrow\rangle = 3.503$.

The energy difference between the two doublets is

$$\Delta E = E_E - E_G = 1.6\text{meV}. \quad (5)$$

### 1.2 Matrix elements for the interaction with an applied magnetic field

Let us now study how an external magnetic field couples to this pair of doublets. The magnetic field acts via the interaction Hamiltonian $H_B(t) = -g_J \mu_B \vec{B}(t) \cdot \vec{J}$, where $g_J = 3/2$ is the Landé factor of the considered Tb$^7F_6$-multiplet and $\mu_B$ is the Bohr magneton. Even though we will discuss general pulses, it is useful to consider the limit of a harmonic driving field $g_J \mu_B \vec{B}(t) = \vec{\mathcal{B}}(t) = \mathcal{B}_0 \vec{e} \exp(-i\omega t) + \text{H.c.}$, where $\mathcal{B}_0 = 1/2 B_0$ is a real amplitude and $\vec{e}$ the polarization vector. For linearly polarized light $\vec{e}$ can be chosen to be real, while circularly polarized light is characterized by a complex $\vec{e}$ in general. For a general pulse shape, $\mathcal{B}_0 \to \mathcal{B}_0(t)$ is a smoothly varying envelope function.

The $J^z$ operator only has matrix elements between states with equal value of $J_z mod 3$ (i.e., between $|G,\uparrow\rangle$ and $|E,\downarrow\rangle$, and between $|G,\downarrow\rangle$ and $|E,\uparrow\rangle$), while the operators $J_\pm = J_x \pm iJ_y$ are purely off-diagonal in the basis of fixed $J_z mod 3$. With the above wavefunctions one finds the following non-zero (real) matrix elements between ground and excited doublets

$$j_z \equiv \langle E,\downarrow |J_z|G,\uparrow\rangle = -\langle E,\uparrow |J_z|G,\downarrow\rangle = 3.029, \quad (6)$$
$$j_\perp \equiv \frac{1}{2}\langle E,\uparrow |J_+|G,\uparrow\rangle = -\frac{1}{2}\langle E,\downarrow |J_-|G,\downarrow\rangle = 2.362. \quad (7)$$

The action of a magnetic field in this pair of doublets is best understood within a rotating wave approximation (RWA). The latter is justified for driving close to the resonance frequency, $\omega \approx \Delta E/\hbar$ and at moderate driving strength ($j_{z,\perp} \mathcal{B}_0 \ll \Delta E = 1.6 meV$), where non-resonant intra-doublet and fast-oscillating inter-doublet couplings can safely be neglected. These conditions are met only approximately by our experimental parameters: the central frequency of the driving pulse is $\hbar\omega = 2.4 \, meV$ and $j_{z,\perp} \mathcal{B}_0 = j_{z,\perp} g_J \mu_B B_{0,p}/2 \approx 0.03 \, meV$ using the experimentally determined magnetic field peak amplitude of $B_{0,p} = 0.25T$ and approximating $j_{z,\perp}$ with a typical value of 3. However, the approximation still gives useful insight into the main effect of the driving. Within the RWA the resulting driving Hamiltonian then reduces to

$$H_B(t) \to H_{RWA}(t) = -\mathcal{B}_0 e^{-i\omega t} \sum_{\alpha,\beta \in \{\uparrow\downarrow\}} M_{\alpha\beta} |E,\alpha\rangle\langle G,\beta| + H.c., \quad (8)$$

where the 2x2 matrix $M_{\alpha\beta}$ is given by

$$M = \sum_{a=x,y,z} e_a \hat{J}_a^{EG} \quad (9)$$

with the inter-doublet matrix elements of the magnetic moment operators given by

$$\begin{aligned}
\hat{J}_x^{EG} &= j_\perp \sigma^z, \\
\hat{J}_y^{EG} &= -ij_\perp, \quad (10) \\
\hat{J}_z^{EG} &= -ij_z \sigma^y.
\end{aligned}$$

where the $\sigma^{x,y,z}$ are the standard Pauli matrices.

**1.3 Temporal evolution in the rotating frame**
We seek the solution to the time-dependent Schrödinger equation $i\partial_t \psi = [H_{CF} + H_{RWA}(t)]\psi$ in the form

$$\psi(t) = \sum_\alpha [e_\alpha(t)|E,\alpha\rangle e^{-iE_E t} + g_\alpha(t)|G,\alpha\rangle e^{-iE_G t}] \quad (11)$$

where at the level of the RWA and for resonant driving $\omega = \Delta E/\hbar$, the coefficients $e_\alpha, g_\alpha$ obey the Schrödinger equation in the rotating frame,

$$\begin{aligned}
i\partial_t e_\alpha(t) &= -\mathcal{B}_0 \sum_\beta M_{\alpha\beta} g_\beta(t), \\
i\partial_t g_\alpha(t) &= -\mathcal{B}_0 \sum_\beta M_{\alpha\beta}^\dagger e_\beta(t). \quad (12)
\end{aligned}$$

Taking a second derivative, one finds a closed differential equation for the ground state amplitude vector $g = \begin{pmatrix} g_\uparrow \\ g_\downarrow \end{pmatrix}$ alone,

$$-\partial_t^2 g(t) = \mathcal{B}_0^2 M^\dagger M g(t), \quad (13)$$

using matrix notation.

**1.4 Splitting of the four-level systems into effective two-level systems**
The Hermitian matrix $M^\dagger M$ is positive definite and diagonalizable. We call $\phi_{1,2}$ its orthonormal eigenvectors, and $d_{1,2}^2$ the corresponding eigenvalues. The explicit form of the matrix $M^\dagger M$ is

$$M^\dagger M = A + \vec{v} \cdot \vec{\sigma}, \quad (14)$$

where

$$\begin{aligned}
A &= |\vec{b}|^2, \\
v_x &= -2\text{Re}(b_z^* b_x), \\
v_y &= 2\text{Re}(b_z^* b_y), \\
v_z &= 2\text{Im}(b_x^* b_y), \quad (15)
\end{aligned}$$

and we have used the notation $b_z \equiv e_z j_z$, $b_{x,y} \equiv e_{x,y} j_\perp$. From this, one finds that $\phi_{1,2}$ are the eigenvectors of $\vec{v} \cdot \vec{\sigma}$ with eigenvalue $\pm|\vec{v}|$. The associated magnetic moments are, respectively,

$$d_{1,2} = \sqrt{A \pm |\vec{v}|}. \quad (16)$$

For linear polarization, the $b_\alpha$ are real, and the magnetic moments evaluate to

$$d_{1,2} = ||b_z| \pm |b_\perp|| = ||e_z j_z| \pm |e_\perp j_\perp||, \quad (17)$$

with $b_\perp = \sqrt{b_x^2 + b_y^2}$, and $e_\perp = \sqrt{e_x^2 + e_y^2}$.

Note that $\phi_i$ only couples to one excited state, $\psi_i$. Therefore, $\{\phi_1, \psi_1\}$ and $\{\phi_2, \psi_2\}$ form two independent effective two-level systems. The $\psi_i$ span an orthonormal basis of the excited doublet space, being the eigenvectors of $MM^\dagger$, for which one finds

$$MM^\dagger = \sigma^x (A - \vec{v} \cdot \vec{\sigma}) \sigma^x. \quad (18)$$

Up to phases its eigenvectors are thus $\psi_{1,2} \propto \sigma^x \phi_{2,1}$.

## 1.5 Driven dynamics of the effective two-level systems

In the basis $\{\phi_i, \psi_i\}$, the full driving Hamiltonian takes the form

$$H_{res} = \begin{pmatrix} 0 & -\mathcal{B}(t)d_1 & 0 & 0 \\ -\mathcal{B}(t)d_1 & \Delta E & 0 & 0 \\ 0 & 0 & 0 & -\mathcal{B}(t)d_2 \\ 0 & 0 & -\mathcal{B}(t)d_2 & \Delta E \end{pmatrix}, \quad (19)$$

where we dropped the non-resonant intra doublet couplings. $d_i$ can be seen as the effective magnetic moment of the two-level systems and $\Omega_{R,i} = \mathcal{B}_0 d_i/\hbar$ is the associated instantaneous Rabi frequency. The $2 \times 2$ Hamiltonian governing the effective TLS generates the unitary, non-dissipative part of the dynamics for the ensemble averaged density matrix $\rho = \overline{|\psi(t)\rangle\langle\psi(t)|}$, $\partial_t \rho = -i/\hbar [H_{res}, \rho]$ of Eqs. (3, 4) in the main text. Defining the resonance frequency $\omega_0 = \Delta E/\hbar$ and the occupation difference $\Delta n(t) = \rho_{00}(t) - \rho_{11}(t)$, we obtain:

$$\frac{d\Delta n(t)}{dt} = \frac{2id\mathcal{B}(t)}{\hbar} (\rho_{10}(t) - \rho_{10}^*(t)); \quad (20)$$

$$\frac{d\rho_{10}(t)}{dt} = -i\omega_0 \rho_{10}(t) + i\frac{d\mathcal{B}(t)}{\hbar} \Delta n(t). \quad (21)$$

Recalling that $\mathcal{B}(t) = g_J \mu_B B(t)$ yields the non-dissipative part of equations (3-4) of the main text, where $|d| \approx j_{typ} = 3$ interpolates between the transverse and longitudinal transition matrix elements, $j_\perp, j_z$.

Given the modest excitation level, one may approximate $\Delta n(t)$ in (22) by $\Delta n_{eq}$, which yields the coherence:

$$\rho_{10}(t) = \frac{i}{\hbar} \Delta n_{eq} d \int_{-\infty}^{t} dt' \, \mathcal{B}(t') e^{\left(-i\omega_0 - \frac{1}{\tau_c}\right)(t-t')}, \quad (22)$$

which is proportional to the occupation difference before excitation $\Delta n_{eq}$.

Note that in the low temperature state (below the Curie-Weiss temperature $|\Theta_{CW}| = 13\,K$), the interactions induce correlations and entanglement among the different Tb ions. However, since we focus on relatively short times as compared to the inverse interaction strength and since we consider only a moderate level of excitation, we neglect those correlations and describe the driving at the level of the single site density matrix. In the main text, the interaction with neighboring sites are summarily accounted for by including a relaxation and dephasing rate (of the order of the Tb-Tb interaction strength) in the evolution equation for the single site density matrix.

## 1.6 Unitary dynamics within RWA approximation

We now return to the RWA ansatz (11), where only the resonant frequency component at $\omega = \pm\omega_0$ of $\mathcal{B}(t)$ is retained. In a thermal ensemble at temperature $k_B T \ll \hbar\omega_0$, the initial wavefunction $\psi(0)$ restricted to a given

Tb site is a random state within the ground doublet, $e_\alpha(t = -\infty) = 0$. Within the RWA the general solution of the resonantly driven Schrödinger equation reads

$$\begin{aligned} g(t) &= C_1 \cos(d_1 F(t)) \phi_1 + C_2 \cos(d_2 F(t)) \phi_2, \\ e(t) &= iC_1 \sin(d_1 F(t)) \psi_1 + iC_2 \sin(d_2 F(t)) \psi_2, \end{aligned} \quad (23)$$

where

$$F(t) = \int_{-\infty}^{t} dt' \mathcal{B}_0(t').$$

Observables pertaining to the subensemble of Tb ions with equally oriented local z-axis have to be averaged with the prescription $\overline{C_i^* C_j} = \delta_{ij}/2$, and $\overline{C_i C_j} = 0$.

### 1.7 Time-evolving magnetization
The time-evolving magnetic moment of the Tb wavefunction is obtained as

$$\begin{aligned} m_\alpha(t) &= g_J \mu_B \langle \psi(t) | J_\alpha | \psi(t) \rangle \\ &= g_J \mu_B \delta_{\alpha,z} [m_G g^\dagger(t) \sigma^z g(t) + m_E e^\dagger(t) \sigma^z e(t)] \\ &\quad + 2 g_J \mu_B \mathrm{Re}[e^{i\omega_0 t} e^\dagger(t) \hat{J}_\alpha^{EG} g(t)]. \end{aligned} \quad (24)$$

Note that once the driving field is switched off, the wavefunctions in the rotating frame, $e(t), g(t)$ do not evolve further and would stay constant in the absence of interactions with other degrees of freedom. The first line in (24) captures the static, time-independent magnetization induced by the driving pulse $\mathcal{B}(t)$, while the second contribution to the magnetization oscillates with the frequency $\omega_0$ and reflects the coherence of the wavefunction, being a superposition of components in the excited and ground state doublet. Interactions with neighboring Tb ions eventually induce dephasing, which degrades the oscillatory term, and relaxation, which makes the static part of the induced magnetization decay.

### 1.8 Oscillatory magnetization induced by linearly polarized light
For linearly polarized light the static part of the magnetization vanishes since the magnetic moments of the light-selected TLS vanish, $\langle \phi_i | J^z | \phi_i \rangle \sim \langle \phi_i | \sigma^z | \phi_i \rangle = \pm \frac{v_z}{|\vec{v}|} = 0$ and likewise for $\langle \psi_i | J^z | \psi_i \rangle$

The vector $\vec{v}$ in (15) is proportional to $(-e_x, e_y, 0)$. The eigenvectors associated to the effective magnetic moments $d_{1,2} = |b_\perp \mp b_z|$ are explicitly

$$\phi_{1,2} = \frac{1}{\sqrt{2}} \begin{pmatrix} \pm(e_x + ie_y)/e_\perp \\ 1 \end{pmatrix}, \quad (25)$$

and

$$\psi_{1,2} = \mathrm{sgn}(b_\perp \mp b_z) \frac{1}{\sqrt{2}} \begin{pmatrix} \pm 1 \\ -(e_x + ie_y)/e_\perp \end{pmatrix}. \quad (26)$$

In this basis, the off-diagonal matrix elements of the magnetic moment operator $\hat{J}^{EG}$ evaluate to

$$\psi_{1,2}^\dagger \begin{pmatrix} \hat{J}_x^{EG} \\ \hat{J}_y^{EG} \\ \hat{J}_z^{EG} \end{pmatrix} \phi_{1,2} = \mathrm{sgn}(b_\perp \mp b_z) \begin{pmatrix} j_\perp e_x / e_\perp \\ j_\perp e_y / e_\perp \\ \mp j_z \end{pmatrix}. \quad (27)$$

The oscillatory part of the magnetization results from Eq. (27) as

$$\begin{aligned} \overline{m_{x,y}^{osc}}(t) &= g_J \mu_B \overline{2\mathrm{Re}[e^{i\omega_0 t} e^\dagger(t) \hat{J}_{x,y}^{EG} g(t)]} \quad (28) \\ &= g_J \mu_B \frac{j_\perp}{2} \frac{e_{x,y}}{e_\perp} \sin(\omega_0 t) \\ &\quad \times (\sin[2(j_\perp e_\perp + j_z e_z) F(t)] + \sin[2(j_\perp e_\perp - j_z e_z) F(t)]), \end{aligned}$$

and

$$\overline{m_z^{osc}}(t) = g_J\mu_B \overline{2Re[e^{i\omega_0 t}e^\dagger(t)\hat{j}_z^{EG}g(t)]} \qquad (29)$$
$$= g_J\mu_B \frac{j_z}{2}\sin(\omega_0 t)$$
$$\times (\sin[2(j_\perp e_\perp + j_z e_z)F(t)] - \sin[2(j_\perp e_\perp - j_z e_z)F(t)]).$$

To include dephasing, we switch to the density matrix formulation, approximating the subspaces of the two light-selected TLS $(\phi_i, \psi_i)_{i=1,2}$ as uncoupled. Let us discuss here the case of relatively weak or short pulses with $F_{max} \ll 1$, as used in the experiment. For the ensemble averaged magnetic moment one finds

$$\overline{m_\alpha(t)} = \frac{1}{2}g_J\mu_B \sum_{i=1,2} 2Re[(\psi_i^\dagger \hat{j}_\alpha^{EG}\phi_i)\rho_{01}^{(i)}], \qquad (30)$$

whereby the off-diagonal term of the density matrix of the TLS $i$, $\rho_{01}^{(i)}$, satisfies Eq. (21) with $d = d_i$. Resorting to the resonant approximation, writing $\vec{\mathcal{B}}(t) = \mathcal{B}_0\vec{e}\exp(-i\omega_0 t) + \text{H.c.}$ and retaining only the positive frequency component, one finds the contributions of the two TLS to add up to yield

$$\overline{m_\alpha(t)} = g_J\mu_B \sum_{i=1,2}(\psi_i^\dagger \hat{j}_\alpha^{EG}\phi_i)d_i$$
$$\times Re[i\exp(-i\omega_0 t)\int_{-\infty}^{t} dt'\mathcal{B}_0(t')e^{-(t-t')/\tau_c}]$$
$$= 2g_J\mu_B j_\alpha^2 e_\alpha \sin(\omega_0 t)\int_{-\infty}^{t} dt'\mathcal{B}_0(t')e^{-(t-t')/\tau_c}, \qquad (31)$$

where we used the notation $j_x = j_y \equiv j_\perp$. Note that in this regime, where the excitation level remains modest, the time dependence of the contributions from the different TLS is the same, and is well captured by a single representative TLS, as was done in the main text.

For a weak pulse and upon neglecting dephasing (setting $\tau_c \to \infty$) this indeed coincides with the linearization of the results (28, 29),

$$\overline{m_{x,y}^{osc}}(t) \approx e_{x,y}(g_J\mu_B j_\perp)^2 \sin(\omega_0 t)\int_{-\infty}^{t} dt' B_0(t'),$$
$$\overline{m_z^{osc}}(t) \approx e_z(g_J\mu_B j_z)^2 \sin(\omega_0 t)\int_{-\infty}^{t} dt' B_0(t'), \qquad (32)$$

where $B_0(t) = 2\mathcal{B}_0(t)/g_J\mu_B$ is the time-dependent linearly polarized magnetic field envelope of the pulse.

### 1.9 Inducing quasi-static magnetization by circularly polarized light

With circularly polarized pulses a quasi-static magnetization can be induced. Using the general solution (20) and (21), the ensemble average of the non-oscillatory magnetization (described by the first line in (24)) can be evaluated as

$$\overline{m_z^{stat}(t)}/g_J\mu_B = \frac{m_G}{2}[\langle\phi_1|\sigma^z|\phi_1\rangle\cos^2(d_1 F(t)) + \langle\phi_2|\sigma^z|\phi_2\rangle\cos^2(d_2 F(t))]$$
$$+ \frac{m_E}{2}[\langle\psi_1|\sigma^z|\psi_1\rangle\sin^2(d_1 F(t)) + \langle\psi_2|\sigma^z|\psi_2\rangle\sin^2(d_2 F(t))]$$
$$= \frac{m_G - m_E}{2}\mu_z[\cos^2(d_1 F(t)) - \cos^2(d_2 F(t))], \qquad (33)$$

where we used (19). This quasi-static magnetization settles to $F(t) \to F_{max}$ after the pulse.

As an example, we consider a CP pulse propagating along the local z-axis of one type of ions, with $B_y = iB_x \equiv iB$, $B_z = 0$. The matrix $M$ takes the form $M = j_\perp B(\sigma^z + 1)$, from which one extracts $\phi_1 = (1,0)$, $d_1 = 2j_\perp = 4.724$; $\phi_2 = (0,1)$, $d_2 = 0$ and $\mu_z = 1$. This results in the induced static magnetization

$$\begin{aligned}
\overline{m_z^{stat,\|}(t)}/g_J\mu_B &= \mu_z \frac{m_E - m_G}{2}(\cos^2(d_1 F(t)) - \cos^2(d_2 F(t))) \\
&= -\frac{m_E - m_G}{2}\sin^2(2j_\perp F(t)) \\
&\approx 0.423\sin^2(4.724 F(t)). \quad (34)
\end{aligned}$$

On average, half of these ions are in the state $\phi_1$ which undergoes Rabi oscillations, a full population conversion leading to a change of magnetization by $m_E - m_G$

For the three differently oriented ions, one finds two non-trivial effective TLS, with effective magnetic moments $d_1' = 4.493$ and $d_2' = 2.919$, respectively, and magnetizations of the associated TLS basis vectors $\mu_z' \equiv \langle\phi_1|\sigma^z|\phi_1\rangle = -0.637$. Injecting this in the above formulae one finds the induced quasi-static magnetization along the ions' respective local z-axes to evaluate to

$$\overline{m_z^{stat,others}(t)}/g_J\mu_B \quad (35)$$
$$= \mu_z' \frac{m_G - m_E}{2}[\cos^2(d_1' F(t)) - \cos^2(d_2' F(t))]$$
$$= 0.2696(\cos^2(4.493 F(t)) - \cos^2(2.919 F(t))). \quad (36)$$

The moments of three different non-aligned ions sum up to a magnetization pointing along $-\vec{e}_z$, and having a magnitude $\overline{m_z^{stat,others}(t)}$.

In TTO a cell of volume $V = (10.3\text{Å})^3/4$ contains four Tb ions, one of each orientation, that we treat as uncorrelated. The total quasistatic magnetic moment induced in such a cell by a CP pulse of action $F_{max}$, oriented along the axis of one of the ions points along the propagation direction of the pulse and has a magnitude

$$m^{tot,stat}/g_J\mu_B = [0.423\sin^2(4.724 F_{max}) \quad (37)$$
$$-0.2696(\cos^2(4.493 F_{max}) - \cos^2(2.919 F_{max}))].$$

For short/weak pulses, the final magnetic moment behaves as $m^{tot,slow}(t)/g_J\mu_B \approx 11.01 F_{max}^2$. The average internal field resulting from this magnetization density follows as

$$B_{int} = \frac{2}{3}\mu_0 \frac{m^{tot,stat}}{V}, \quad (38)$$

which can reach of the order of $20 mT$ in Tb$_2$Ti$_2$O$_7$.

Due to interactions between the ions, the magnetization will, however, decay back to zero (thermal equilibrium) over a time scale of the order of $\tau$ via diffusion of the excitations. On the other hand the global excitation level (total energy) of the ions relaxes much more slowly, necessitating coupling to the phonon bath.

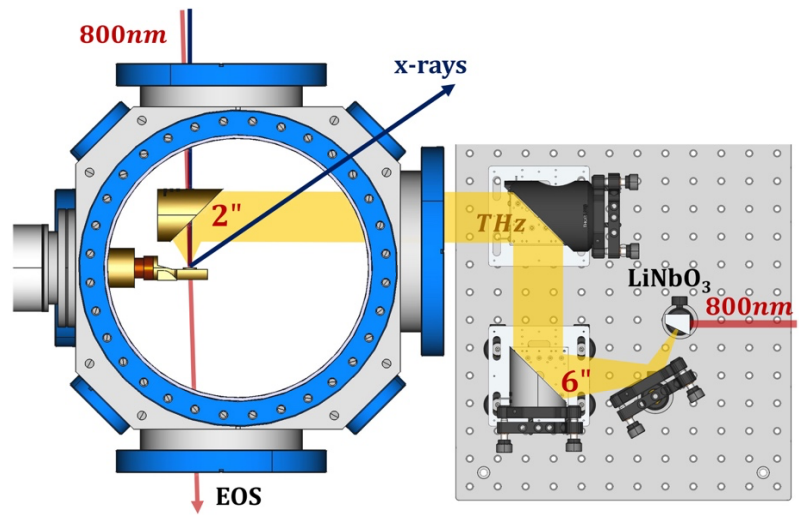

**Fig. S1. Experimental geometry.** The THz pulses (yellow) were generated outside the chamber and passed through a window into the vacuum chamber. They were focused with a 2" focal distance parabolic mirror in vacuum onto the sample and overlapped with the x-ray probe pulses (dark blue) on the sam,ple. The THz electromagnetic field was characterized by electrooptic sampling (EOS) using 800nm pulses (red) outside the chamber.

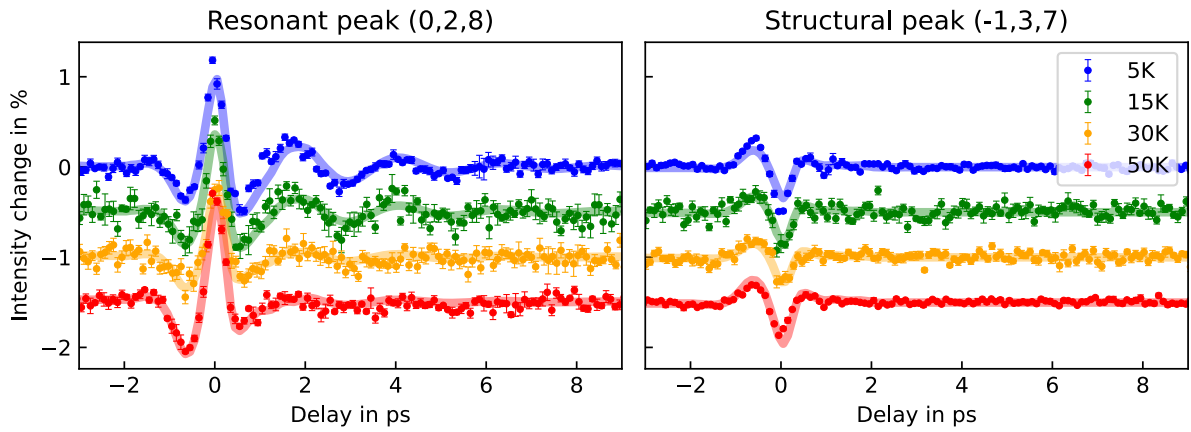

**Fig. S2. Temperature dependence of the intensity change signal of the structural and resonant peaks.** Intensity changes induced by the same THz excitation are shown at different temperatures for the resonant (028) peak (left side) and the structurally allowed (-137) peak.

**Table S1.** Summary of the 4 unique Tb ion positions and local quantization axis, as well as the induced magnetic moment $\vec{m_i}$ by the excitation on each site $i$ with: $m_{\perp,z} = (g_J \mu_B j_{\perp,z})^2 \sin(\omega_0 t) \int_{-\infty}^{t} dt' B(t') e^{-(t-t')/\tau_c}$. The last column denotes their phase factors for the $(hkl) = (028)$ peak.

| Tb | Position $xyz$ | $x_{local,i}$ | $y_{local,i}$ | $z_{local,i}$ | Local magnetic moment $\vec{m_i}$ on Tb site $i$ | $e^{2\pi i(hx+ky+lz)}$ |
|---|---|---|---|---|---|---|
| 1 | (0, 0, 0) | [1 -1 0] | [1 1 -2] | [1 1 1] | $\left(\frac{m_\perp}{\sqrt{2}}, -\frac{m_\perp}{\sqrt{2}}, 0\right)$ | 1 |
| 2 | (3/4, 1/4, 1/2) | [-1 1 0] | [-1 -1 -2] | [-1 -1 1] | $-\left(\frac{m_\perp}{\sqrt{2}}, -\frac{m_\perp}{\sqrt{2}}, 0\right)$ | -1 |
| 3 | (1/4, 1/2, 3/4) | [-1-1 0] | [-1 1 2] | [-1 1 -1] | $\left(-\frac{0.58m_\perp}{\sqrt{6}} - \frac{0.82m_z}{\sqrt{3}}, \frac{0.58m_\perp}{\sqrt{6}} + \frac{0.82m_z}{\sqrt{3}}, \frac{1.16m_\perp}{\sqrt{6}} - \frac{0.82m_z}{\sqrt{3}}\right)$ | 1 |
| 4 | (1/2, 3/4, 1/4) | [1 1 0] | [1 -1 2] | [1 -1 -1] | $\left(\frac{0.58m_\perp}{\sqrt{6}} + \frac{0.82m_z}{\sqrt{3}}, -\frac{0.58m_\perp}{\sqrt{6}} - \frac{0.82m_z}{\sqrt{3}}, \frac{1.16m_\perp}{\sqrt{6}} - \frac{0.82m_z}{\sqrt{3}}\right)$ | -1 |

**Table S2.** Parameters used for the simulation of the time-dependent diffraction intensities of the structural $(\bar{1}37)$ and resonant $(028)$ Bragg peaks.

| $a_s$ in %/(MV/cm) | $a_{s,o}$ in %/(MV/cm) | $a_m$ in % | $\nu_0$ in $THz$ | $\tau_c$ in $ps$ | $\tau$ in $ps$ | $\Delta n_{eq}$ |
|---|---|---|---|---|---|---|
| $0.66 \pm 0.04$ | $-1.1 \pm 0.07$ | $1.5 \pm 0.3$ | $0.44 \pm 0.02$ | $2.4 \pm 0.4$ | $\approx 4$ | 1 |